\documentclass[twoside,twocolumn,9pt]{article}
\usepackage{extsizes}
\usepackage[super,sort&compress,comma]{natbib} 
\usepackage[version=3]{mhchem}
\usepackage[left=1.5cm, right=1.5cm, top=1.785cm, bottom=2.0cm]{geometry}
\usepackage{balance}
\usepackage{widetext}
\usepackage{times,mathptmx}
\usepackage{sectsty}
\usepackage{graphicx} 
\usepackage{lastpage}
\usepackage[format=plain,justification=raggedright,singlelinecheck=false,font={stretch=1.125,small,sf},labelfont=bf,labelsep=space]{caption}
\usepackage{float}
\usepackage{fancyhdr}
\usepackage{fnpos}
\usepackage[english]{babel}
\usepackage{array}
\usepackage{droidsans}
\usepackage{charter}
\usepackage[T1]{fontenc}
\usepackage[usenames,dvipsnames]{xcolor}
\usepackage{setspace}
\usepackage[compact]{titlesec}

\newcommand{\br}{{\bf r}}
\newcommand{\brp}{{{\bf r}}'}
\newcommand{\rp}{{{ r}}'}

\newcommand{\gj}[6]{ \begin{pmatrix}
  #1 & #2 & #3 \\
  #4 & #5 & #6 
 \end{pmatrix}}

\newcommand*{\citen}[1]{%
  \begingroup
    \romannumeral-`\x 
    \setcitestyle{numbers}%
    \cite{#1}%
  \endgroup   
}

\newcommand{\eq}[1]{eqn~(\ref{#1})}
\newcommand{\Eq}[1]{Eqn~(\ref{#1})}

\definecolor{cream}{RGB}{222,217,201}

\DeclareMathAlphabet{\mathcal}{OMS}{cmsy}{m}{n}

\begin{document}

\pagestyle{fancy}
\thispagestyle{plain}

\makeFNbottom
\makeatletter
\renewcommand\LARGE{\@setfontsize\LARGE{15pt}{17}}
\renewcommand\Large{\@setfontsize\Large{12pt}{14}}
\renewcommand\large{\@setfontsize\large{10pt}{12}}
\renewcommand\footnotesize{\@setfontsize\footnotesize{7pt}{10}}
\makeatother

\renewcommand{\thefootnote}{\fnsymbol{footnote}}
\renewcommand\footnoterule{\vspace*{1pt}%
\color{cream}\hrule width 3.5in height 0.4pt \color{black}\vspace*{5pt}} 
\setcounter{secnumdepth}{5}

\makeatletter 
\renewcommand\@biblabel[1]{#1}            
\renewcommand\@makefntext[1]%
{\noindent\makebox[0pt][r]{\@thefnmark\,}#1}
\makeatother 
\renewcommand{\figurename}{\small{Fig.}~}
\sectionfont{\sffamily\Large}
\subsectionfont{\normalsize}
\subsubsectionfont{\bf}
\setstretch{1.125} 
\setlength{\skip\footins}{0.8cm}
\setlength{\footnotesep}{0.25cm}
\setlength{\jot}{10pt}
\titlespacing*{\section}{0pt}{4pt}{4pt}
\titlespacing*{\subsection}{0pt}{15pt}{1pt}

\makeatletter 
\newlength{\figrulesep} 
\setlength{\figrulesep}{0.5\textfloatsep} 

\newcommand{\topfigrule}{\vspace*{-1pt}%
\noindent{\color{cream}\rule[-\figrulesep]{\columnwidth}{1.5pt}} }

\newcommand{\botfigrule}{\vspace*{-2pt}%
\noindent{\color{cream}\rule[\figrulesep]{\columnwidth}{1.5pt}} }

\newcommand{\dblfigrule}{\vspace*{-1pt}%
\noindent{\color{cream}\rule[-\figrulesep]{\textwidth}{1.5pt}} }

\makeatother

\twocolumn[
  \begin{@twocolumnfalse}
\sffamily
\centering
\begin{tabular}{m{1cm} m{15cm} m{1cm}}

& \noindent\LARGE{\textbf{Interaction of  molecular nitrogen with free-electron laser\linebreak radiation }} \\
\vspace{0.3cm} & \vspace{0.3cm} &  \\

& \noindent\large{H. I. B. Banks\textit{$^{a}$}, D. A. Little \textit{$^{a}$}, J. Tennyson\textit{$^{a}$} and A. Emmanouilidou\textit{$^{a}$}} & \\

\vspace{0.3cm} & \vspace{0.3cm} &  \\

& \noindent\normalsize{
We compute molecular continuum orbitals in the single center expansion scheme. We then employ these orbitals to obtain molecular Auger rates and single-photon ionization cross sections to study the interaction of $\mathrm{N_2}$ with free-electron laser (FEL) pulses. The nuclei are kept fixed. We formulate rate equations for the energetically allowed molecular and atomic transitions  and we account for dissociation through additional terms in the rate equations. Solving these equations for different parameters of the FEL pulse, allows us  to identify the most efficient parameters of the FEL pulse for obtaining the highest contribution of double core hole (DCH) states in the final atomic ion fragments. Finally we identify the contribution of DCH states in the electron spectra and show that the DCH state contribution is more easily identified in the photo-ionization rather than the Auger transitions.} & \\

\end{tabular}

 \end{@twocolumnfalse} \vspace{0.6cm}

  ]

\renewcommand*\rmdefault{bch}\normalfont\upshape
\rmfamily
\section*{}
\vspace{-1cm}


\footnotetext{\textit{$^{a}$~Department of Physics and Astronomy, University College London, Gower Street, London WC1E 6BT, United Kingdom. E-mail: a.emmanouilidou@ucl.ac.uk}}


\section{Introduction}

The development of x-ray free-electron lasers (FELs)\cite{FELhistory} has introduced new tools for imaging and exploring novel states of atoms and molecules.\cite{Marangos2011,Ullrich2012} Potential applications of FELs range from imaging biomolecules\cite{Schlichting2012,Neutze, Redecke227} to accurate modeling of laboratory and astrophysical plasmas. FEL driven processes in atoms and molecules include single-photon ionization and Auger processes. Sequential single-photon ionization processes occurring on a time scale that is faster than Auger decays lead to the formation of inner-shell holes in atoms and molecules. In an Auger process a valence electron drops in to fill a core hole; the energy released allows another valence electron to escape to the continuum. The formation of double core hole (DCH) states in molecules is of particular interest for chemical analysis \cite{Cederbaum1986,Tashiro2010}. The energy required to remove a core electron depends upon the chemical environment of the site the electron is removed from rendering DCHs  a sensitive spectroscopic tool for chemical analysis.


To understand the formation and detection of single core hole (SCH) and DCH states in molecules, one must explore the interplay of  Auger  and photo-ionization processes. There has been a significant amount of work on calculating the Auger rates and photo-ionization cross sections in atoms.\cite{Pulkkinen1995,Wallis,Lablanquie2007,Bhalla,Son2012,Makris2009} The work on computing these rates for molecules is significantly less. The reason is that molecules do not have  spherical symmetry and thus computing the molecular continuum orbital of the escaping electron is a complex task. This in turn hinders the computation of the Auger rates and the single-photon ionization cross sections in molecules. 

Previous molecular studies with FEL radiation include models where the molecule is treated as a combination of atoms. Then, in these models, the Auger rates and the photo-ionization cross sections are computed for atomic transitions.\cite{Buth2012,LiuBerrah} These atomic rates  are then used to setup rate equations to describe molecular interactions with FEL radiation. In some instances dissociation  is accounted for through additional terms in the rate equations.\cite{Buth2012,LiuBerrah} For high photon energy FEL pulses interacting with $\mathrm{N_{2}}$, these models have been used to compute the yields of the final atomic ion fragments as well as the contribution of the DCH molecular  states in the yields of the final atomic ions.\cite{Buth2012,LiuBerrah} Very recently, new methods have been developed to describe molecular states with multiple holes and to compute molecular transitions following interaction  with FEL radiation.\cite{Hao2015, Jurek2016} These new methods have been employed to compute the yields of the final molecular ion states as well as the contribution of SCH and DCH states 
in   water for fixed nuclei.\cite{Inhester2016} The calculations in these studies were performed with atomic continuum orbitals rather than molecular ones. The use of atomic continuum orbitals is a good approximation when these models are employed to study molecular interactions with high photon energy FEL pulses. 

$\mathrm{N_2}$ interacting with FEL pulses  has been the subject of many experimental studies.\cite{Fang, Cryan2012, Ueda2009} In these studies the yields of the final atomic ion states and the formation of molecular DCHs are investigated. In this work, we study the interaction of the  $\mathrm{N_2}$ diatomic molecule  with FEL radiation. 
To do so, we assume that the nuclei are fixed, an assumption also made in previous studies.\cite{Buth2012, LiuBerrah, Inhester2016} Very importantly, we compute the molecular continuum orbitals. We then employ  these orbitals to compute the Auger rates and the single-photon ionization cross sections for all molecular transitions that are energetically accessible.    This way our studies are not restricted to high photon FEL pulses. Specifically, we investigate the interaction of a  525 eV and a 1100 eV FEL pulse with $\mathrm{N_2}$. These photon energies are sufficient to create  three inner-shell holes through sequential single-photon absorptions and multiple valence holes in the ground state of $\mathrm{N_2}$.  Moreover, as we show in the section concerning electron spectra, for a 525 eV FEL pulse some of the electrons ionize with very small energies. These small energies  render necessary the use of molecular continuum orbitals, as is done in the current work. We compute the Auger and the single-photon ionization processes for the molecular transitions allowed, thus improving over previous studies that consider only atomic transitions.\cite{LiuBerrah} We  then set  up rate equations for the allowed molecular and atomic transitions  and account for dissociation through additional terms in the rate equations. We also investigate the dependence of the final  molecular and atomic ion fragments on the intensity and pulse duration of the FEL pulse. Moreover, we compute all energetically accessible pathways and can thus determine the contribution of the DCH molecular states in the final atomic ion states as well as in the electron spectra. 
Finally, we investigate whether photo-ionization or Auger transitions in the electron spectra are more effective in detecting  the formation of DCH molecular states.


\section{Ion Yields and Pathways}

We study the response of $\mathrm{N_{2}}$ to  FEL pulses of photon energies 525 eV and 1100 eV. We do so by formulating and solving a set of rate equations for the time dependent population of the ion states.

\subsection{Rate Equations}

\begin{figure}
\begin{center}
\includegraphics[width=1\linewidth]{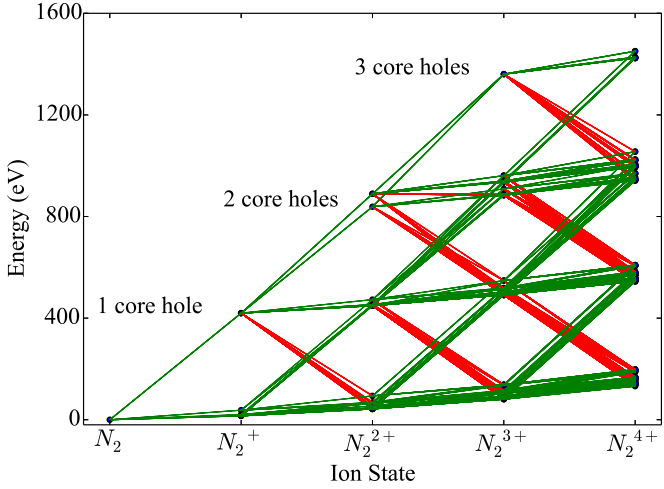}
\caption{Ionization pathways between different electronic configurations of $\mathrm{N_{2}}$ up to $\mathrm{N_{2}^{4+}}$, accessible with sequential single-photon ($\mathrm{\hbar \omega=525 eV}$) absorptions and Auger decays. The green and red lines indicate photo-ionization  and Auger transitions, respectively}
\label{fig:N2_energylevels}
\end{center}
\end{figure}

We construct rate equations for each energetically accessible state of molecular nitrogen. Each molecular state  is denoted by  its electronic configuration ($\mathrm{1\sigma_g^{a},1\sigma_u^{b},2\sigma_g^{c},2\sigma_u^{d},1\pi_{ux}^{e},1\pi_{uy}^{f}, 3\sigma_g^{g}}$) with $\mathrm{a,b,c,d,e,f,g}$  the number of electrons occupying a molecular orbital. Each occupation number is equal to  0, 1 or 2.  2 corresponds  to the maximum occupancy of a molecular orbital of  two electrons   with spin up and  down. Atomic units are used in this work, unless otherwise stated.  In Fig. \ref{fig:N2_energylevels}, accounting for states up to $\mathrm{N_2^{4+}}$, we illustrate the photo-ionization and Auger transitions between  molecular states  that are accessible due to the  interaction of $\mathrm{N_{2}}$ with a 525 eV laser pulse.  To create a core hole, a minimum photon energy of 420 eV is required. The transitions in Fig. \ref{fig:N2_energylevels} were calculated for the ground state equilibrium distance of the nuclei for $\mathrm{N_2}$. This was done by employing  Molpro\cite{Molpro} and performing a Hartree Fock calculation using  correlation-consistent polarized triple-zeta (cc-pVTZ) basis set. 

We assume that the nuclei are fixed at the equilibrium distance of $\mathrm{N_{2}}$ of 2.08 a.u.. To model the fragmentation of $\mathrm{N_2}$ that leads to the formation of atomic ions, we include in the rate equations terms accounting for dissociation. Specifically,  we assume that instantaneous dissociation, i.e.  a very large dissociation rate, takes place from  the $\mathrm{N_2^{4+}}$ and $\mathrm{N_2^{3+}}$  molecular states with no core holes. Dissociation of  $\mathrm{N_2^{3+}}$ and $\mathrm{N_2^{4+}}$ leads  to   $\mathrm{N^{+}+N^{2+}}$ and $\mathrm{N^{2+}+N^{2+}}$, respectively. We further assume that 
 all $\mathrm{N_2^{2+}}$ molecular states  dissociate with a lifetime of ~100 fs.\cite{Beylerian2004} $\mathrm{N_2^{2+}}$ dissociation leads  to  $\mathrm{N^{+}+N^{+}}$  and $\mathrm{N+N^{2+}}$ with probabilities $\mathrm{74\%}$ and $\mathrm{26\%}$,\cite{Beylerian2004}  respectively. The above dissociation processes are similar to the dissociation processes taken into account in ref. \citen{LiuBerrah}.  In the rate equations fluorescence is not accounted for since the corresponding rates are much smaller compared to Auger and single-photon processes.\cite{Coville1991}
In the rate equations we also account for the  states  of atomic nitrogen that are accessible through molecular fragmentation and through single-photon ionization and Auger processes occurring in  the atomic ion states. Each atomic state  is denoted by ($\mathrm{1s^a,2s^b,2p^c}$) with a, b and c the occupation numbers of the atomic orbitals. 

 The rate equations describing the population $\mathrm{\mathcal{I}_j(t)}$ of a molecular ion state j take the form

\begin{flalign}
\mathrm{\frac{d}{dt}\mathcal{I}_j(t)} &\mathrm{= \sum_i  \left(\sigma_{i\rightarrow j} J(t)+ \Gamma_{i\rightarrow j} \right) \mathcal{I}_i(t) } \label{N_2RE}
 && \\\nonumber
 &\mathrm{ -  \sum_k \left( \sigma_{j\rightarrow k} J(t) + \Gamma_{j\rightarrow k} \right)  \mathcal{I}_j(t)-  \sum_{n} \kappa_{j\rightarrow n,p}  \mathcal{I}_j(t), }
&& \\\nonumber
 \mathrm{\frac{d}{dt} \mathcal{A}_{i\rightarrow j}} &\mathrm{= \Gamma_{i\rightarrow j} \mathcal{I}_i(t),} &&\\\nonumber
 \mathrm{\frac{d}{dt}\mathcal{P}_{i\rightarrow j}} &\mathrm{= \sigma_{i\rightarrow j} J(t) \mathcal{I}_i(t).} && \nonumber
\end{flalign}

 $\mathrm{\sigma_{i\rightarrow j}}$ and $\mathrm{\Gamma_{i\rightarrow j}}$ are the molecular single-photon absorption cross section and Auger decay rate, respectively, from the initial molecular state i to the final molecular state j. J(t) is the photon flux at time t. 
 The temporal form of the FEL flux is modeled with a Gaussian function given by\cite{Rohringer2007}
 \begin{flalign}
\mathrm{J(t) = 1.554\times 10^{-16} \frac{I_0\textrm{[W cm}^{-2}]}{\hbar\omega\textrm{[eV]}} \exp\left\{ -4\ln2 \left(\frac{t}{\tau_X}\right)^2 \right\}}&&
\label{Jt}
\end{flalign}
with $\mathrm{\tau_\chi}$ the full width at half maximum (FWHM) and $\mathrm{I_0}$ the peak intensity. The molecular states i, j and k have charges q-1, q and q+1, respectively. 
 $\mathrm{\kappa_{j\rightarrow n,p}}$ denotes the dissociation rate from the initial molecular state j with charge q to the  final atomic states n and p. The atomic states n and p have total charge equal to q. For the dissociation cases currently considered, for each atomic final fragment n  there is only one atomic final fragment p.  The first term in \eq{N_2RE} accounts for the formation of the molecular state j through the single-photon ionization and Auger decay of the molecular state i. The second term  in \eq{N_2RE} accounts for the depletion of the molecular state j by transitioning to a molecular state k through single-photon ionization and Auger decay. 
 The third term accounts for the depopulation of the molecular state j through dissociation to the atomic states n and p.  
 These rate equations, \eq{N_2RE}, are used to calculate the molecular  ion yields.
 In addition, in \eq{N_2RE}, we  solve for the Auger  $\mathrm{\mathcal{A}_{i\rightarrow j}}$ and the photo-ionization  $\mathrm{\mathcal{P}_{i\rightarrow j}}$ yield from an initial molecular state $\mathrm{i}$ with charge q-1 to a final molecular state $\mathrm{j}$ with charge q. These yields provide the probability for observing an electron with energy corresponding to the molecular transition $\mathrm{i\rightarrow j}$. We use these yields to  describe the molecular transitions in the electron spectra produced by the interaction of  $\mathrm{N_{2}}$ with an FEL pulse. The electron spectra are presented   later in the paper.

The rate equations describing the populations of an atomic state n take the form
\begin{flalign}
\mathrm{\frac{d}{dt}\mathcal{I}_n(t)} &\mathrm{= \sum_m  \left(\sigma_{m\rightarrow n} J(t)+ \Gamma_{m\rightarrow n} \right) \mathcal{I}_m(t)}\label{N_2REatom}&&\\\nonumber
 &\mathrm{ -  \sum_o \left( \sigma_{n\rightarrow o} J(t) + \Gamma_{n\rightarrow o} \right)  \mathcal{I}_n(t)+\sum_{j} \frac{\kappa_{j\rightarrow n,p}}{2-\delta_{n,p}}  \mathcal{I}_j(t),}
 && \\\nonumber
 \mathrm{\frac{d}{dt} \mathcal{A}_{m\rightarrow n}} &\mathrm{= \Gamma_{m\rightarrow n} \mathcal{I}_m(t),}&&\\\nonumber
 \mathrm{\frac{d}{dt}\mathcal{P}_{m\rightarrow n}} &\mathrm{= \sigma_{m\rightarrow n} J(t) \mathcal{I}_m(t),} &&
\end{flalign}
where the indices n, m and o refer to atomic states with charges, q, q-1 and q+1, respectively, while j refers to molecular states. The first term in \eq{N_2REatom} accounts for the formation of the atomic  state n through the single-photon ionization and Auger decay of the atomic state m. The second term  in \eq{N_2REatom} accounts for the depletion of the atomic state n by transitioning to an atomic state o through single-photon ionization and Auger decay. 
 The third term in \eq{N_2REatom}  accounts for the population of state n as a result of dissociative transitions from a molecular state j. The factor $\mathrm{\tfrac{1}{2-\delta_{n,p}}}$ conserves the population transfer from the molecular state j. Namely, if the molecular state results in the same two atomic fragments the factor is equal to 1. If fragmentation results  in two different atomic fragments the factor is equal to   $\mathrm{\tfrac{1}{2}}$, since a rate equation is formulated for each atomic fragment separately. As in the molecular rate equations,  in \eq{N_2REatom}, we  solve for the Auger  $\mathrm{\mathcal{A}_{m\rightarrow n}}$ and the photo-ionization  $\mathrm{\mathcal{P}_{m\rightarrow n}}$ yields from an initial atomic state $\mathrm{i}$ with charge q-1 to a final atomic state $\mathrm{j}$ with charge q. These yields provide the probability for observing an electron with energy corresponding to the atomic transition $\mathrm{m\rightarrow n}$. We use these yields to  describe the atomic transitions in the electron spectra produced by the interaction of  $\mathrm{N_{2}}$ with an FEL pulse. 
We note that  \eq{N_2RE} and  \eq{N_2REatom} are solved simultaneously.  We obtain the molecular and atomic ion yields long after the end of the laser pulse.

It is also of interest to compute the population  transfer through  a specific pathway  $\mathrm{i\rightarrow j\rightarrow k}$ where the initial state is i and the final one that is k  is reached through the state j.  The three types of relevant rate equations for computing the pathway populations are given by eqns (\ref{eq:pathMol2Mol}-\ref{eq:pathAtom2Atom})

 \begin{flalign}\label{eq:pathMol2Mol}
\mathrm{\frac{d}{dt} \mathcal{I}_{i\rightarrow j\rightarrow k}(t)}
&= \mathrm{(\sigma_{j\rightarrow k}J(t)+\Gamma_{j\rightarrow k})\mathcal{I}_{i\rightarrow j}(t)}&& \\ \nonumber 
&\mathrm{- \sum_{l}(\sigma_{k\rightarrow l}J(t)+\Gamma_{k\rightarrow l})\mathcal{I}_{i\rightarrow j\rightarrow k}(t)-  \sum_{n} \kappa_{k\rightarrow n,p}  \mathcal{I}_{i\rightarrow j\rightarrow k}(t),}&&
\end{flalign}
\begin{flalign}\label{eq:pathMol2Atom}
\mathrm{\frac{d}{dt} \mathcal{I}_{i\rightarrow j\rightarrow n}(t)}&=\mathrm{\frac{\kappa_{j\rightarrow n,p}}{2-\delta_{n,p}}  \mathcal{I}_{i\rightarrow j}(t) }&&\\\nonumber
&-\mathrm{\sum_{o}(\sigma_{n\rightarrow o}J(t)+\Gamma_{n\rightarrow o})\mathcal{I}_{i\rightarrow j\rightarrow n}(t),}&&
\end{flalign}
\begin{flalign}\label{eq:pathAtom2Atom}
\mathrm{\frac{d}{dt} \mathcal{I}_{i\rightarrow m\rightarrow n}(t)}&=\mathrm{\left(\sigma_{m\rightarrow n} J(t)+ \Gamma_{m\rightarrow n} \right) \mathcal{I}_{i\rightarrow m}(t)}&&\\
&-\mathrm{\sum_{o}(\sigma_{n\rightarrow o}J(t)+\Gamma_{n\rightarrow o})\mathcal{I}_{i\rightarrow m\rightarrow n}(t).}&&\nonumber
 \end{flalign}
The indices i, j, k and l refer to molecular states whereas the indices m, n and o refer to atomic states. \Eq{eq:pathMol2Mol} computes molecular pathway populations, \eq{eq:pathMol2Atom} computes pathway populations where the final state is an atomic one, but the previous states were molecular. Pathway populations where the final and the previous states are atomic ones are computed using  \eq{eq:pathAtom2Atom}. Solving eqns (\ref{eq:pathMol2Mol}-\ref{eq:pathAtom2Atom}), allows us to register all energetically-allowed pathways.  Each pathway starts from the ground state of $\mathrm{N_{2}}$ and ends at an accessible atomic or molecular ion state. Obtaining the pathway populations allows to determine the percentage of final ion states that were formed through pathways involving a molecular states with only a single  or   a double core hole. These results are presented later in the paper.

\subsection{Electron Continuum molecular orbitals }

An advantage of the rate equations formulated in the previous section is that we compute the single-photon ionization cross sections and the Auger rates using the continuum wave functions of the molecular orbitals.  We compute these continuum molecular orbitals by following the formulation in ref. \citen{demekhin11}. In what follows, we briefly outline the steps we follow to compute the continuum molecular orbitals.  The first step in the derivation involves the Hartree-Fock equations \cite{BranJoa} given by
\begin{widetext}
\begin{flalign}
\label{eqn:schr1}
\underbrace{\mathrm{-\frac{1}{2} {\nabla}^2 \phi_{\epsilon}(\br)}}_{\text{Kinetic energy}} + \underbrace{\mathrm{\sum_{n}^{nuc.} \frac{-Z_n}{|\br-{{\bf R}_n}|}\phi_{\epsilon}(\br)}}_{\text{Electron-nuclei}}
+\underbrace{\mathrm{\sum_i^{orb.} a_i \int d\brp \frac{\phi^*_{i}(\brp)\phi_{i}(\brp)}{|\br - \brp|} \phi_{\epsilon}(\br)}}_{\text{Direct interaction}} - \underbrace{\mathrm{\sum_{i}^{orb.} b_i \int d\brp \frac{\phi^{*}_{i}(\brp)\phi_{\epsilon}(\brp)}{|\br - \brp|}\phi_{i}(\br)}}_{\text{Exchange interaction}} \mathrm{= \epsilon\phi_{\epsilon}(\br).}&&
\end{flalign}  
\end{widetext}
\noindent The index $\mathrm{\epsilon}$ denotes a continuum molecular orbital with  $\epsilon>0$ the energy of the ionizing electron. The index i denotes  bound molecular orbitals, where $\mathrm{a_i}$ and $\mathrm{b_i}$ are coefficients associated with the orbital i. These coefficients are derived in Appendix \ref{App:aAndb}.  $\mathrm{{\bf R}_n}$ denotes the position of nucleus n. The electron-nuclei term is the Coulomb interaction of the continuum electron with each one of the nuclei. The direct and exchange terms arise from the Coulomb interaction of the continuum electron with the bound electrons. To simplify the numerical integrations involved in \eq{eqn:schr1}, the bound and continuum orbital wave functions are expressed using  the single-centre expansion (SCE).\cite{demekhin11} This approximation allows for the angular part of the integrations in \eq{eqn:schr1} to be obtained analytically.
According to the SCE the wave function of the molecular orbital i is expressed as 
\begin{flalign}
\mathrm{\phi_i(\textbf{r})=\sum_{lm}\frac{P^i_{lm}(r)Y_{lm}(\theta,\phi)}{r},}&&
\label{SCE}
\end{flalign}
with $\mathrm{\textbf{r}=(r,\theta,\phi)}$ denoting the position of the electron. For continuum orbitals, i is replaced by the energy of the ionizing electron $\mathrm{\epsilon}$. $\mathrm{Y_{lm}}$ is a spherical harmonic with quantum numbers l and m. $\mathrm{P^i_{lm}(r)}$ are single centre expansion coefficients for the orbital i. 
Substituting   \eq{SCE} in \eq{eqn:schr1} and then multiplying  by $\mathrm{Y^{*}_{lm}}$  and  integrating over the angular part results in

\begin{widetext}
\begin{flalign}
\label{eqn:mat1}
 \mathrm{\sum_{{l}'{m}'} \left[ \left( -\frac{d^2}{dr^2} + \frac{l(l+1)}{r^2} -2\epsilon \right)\delta_{l{l}'}\delta_{m{m}'} +  2V^{ne}_{lm,{l}'{m}'}(r) + 2J^{ee}_{lm,{l}'{m}'}(r)   \right]P^\epsilon_{{l}'{m}'}(r) + 2X_{lm}[\bar{P}^\epsilon](r) = 0.}&&
\end{flalign}
\end{widetext}
\noindent $\mathrm{V^{ne}_{lm,{l}'{m}'}(r)}$ is the electron-nuclei interaction, $\mathrm{J^{ee}_{lm,{l}'{m}'}(r)}$ is the  direct  interaction term, and  $\mathrm{X_{lm}[\bar{P}_{\epsilon}](r)}$ is the exchange interaction term. In \eq{eqn:mat1}, we are solving for $\mathrm{P^\epsilon_{{l}'{m}'}(r)}$, which are the coefficients in the SCE of the continuum wavefunction. $\mathrm{\bar{P}^{\epsilon}}$ is the vector form of $\mathrm{P^\epsilon_{{l}'{m}'}(r)}$. 
The electron-nuclei interaction $\mathrm{V^{ne}_{lm,{l}'{m}'}(r)}$ is given by
\begin{flalign}
\label{eqn:gen}
&\mathrm{V^{ne}_{lm,{l}' {m}'}(r) =  \sum_{n}^{nuc. \rm} - Z_n (-1)^m  \sqrt{(2l + 1)(2{l}' +1)}}&& \\\nonumber
&\times\mathrm{\sum_{k q}} \begin{pmatrix}
\mathrm{l} & \mathrm{k} & \mathrm{{l}'} \\ 
\mathrm{0} & \mathrm{0} & \mathrm{0}\end{pmatrix}
\begin{pmatrix}
\mathrm{l} & \mathrm{k} & \mathrm{{l}'} \\ 
\mathrm{-m} & \mathrm{q} & \mathrm{{m}'}
\end{pmatrix} \mathrm{\sqrt{\frac{4 \pi}{2k+1}} Y^*_{k q}(\theta_n,\phi_n)  \frac{r^{k}_{<}}{r^{k+1}_{>}},}&&
\end{flalign}
with the angular integration expressed in terms of Wigner 3j-symbols;\cite{AtoSpec} $\mathrm{r_<=min(r,R_n)}$ and $\mathrm{r_>=max(r,R_n)}$. The direct interaction $\mathrm{J_{lm,l'm'}(r)}$ is given by 
\begin{flalign}
\label{eqn:Jdev3}
&\mathrm{J_{lm,{l}' {m}'}(r) = \sum_{i} a_i \sum_{\substack{\mathrm{l_2 m_2,l_3 m_3,}\\\mathrm{k q}}}  \sqrt{(2l + 1)(2{l}' + 1) (2l_2 + 1)(2l_3 + 1)}}&&\nonumber\\\nonumber
&\times\begin{pmatrix}\mathrm{l_2} & \mathrm{k} & \mathrm{l_3} \\ \mathrm{0} & \mathrm{0} & \mathrm{0}\end{pmatrix}
\begin{pmatrix}\mathrm{l_2} & \mathrm{k} & \mathrm{l_3} \\ \mathrm{-m_2} & \mathrm{q}  & \mathrm{m_3}\end{pmatrix} 
\begin{pmatrix}\mathrm{{l}'} & \mathrm{k} & \mathrm{l} \\ \mathrm{0} & \mathrm{0} & \mathrm{0}\end{pmatrix}
\begin{pmatrix} \mathrm{{l}'} & \mathrm{k}  & \mathrm{l}  \\ \mathrm{-{m}'} & \mathrm{q} & \mathrm{m}\end{pmatrix}&&\\
 &\mathrm{\times{(-1)}^{m_2+{m}'}\int_0^\infty \frac{r^{k}_{<}}{r^{k+1}_{>}}  P^{i*}_{l_2m_2}({r}')P^i_{l_3 m_3}({r}') d{r}',}
 \end{flalign}
where $\mathrm{r_<=min(r,r')}$ and $\mathrm{r_>=max(r,r')}$ and, $\mathrm{l_2,m_2}$ and $\mathrm{l_3,m_3}$ refer to the orbital i. The exchange interaction can be cast as a functional of the SCE coefficients of the continuum electron orbital as follows
\begin{flalign}
&\mathrm{X_{lm}[\bar{P}^{\epsilon}](r)  =  \sum_{{l}'{m}'} \sum_{i}^{orb.} b_i  \sum_{\substack{\mathrm{l_2 m_2,l_3 m_3,}\\\mathrm{k q}}} \sqrt{(2l + 1)(2{l}' + 1) (2l_2 + 1)(2l_3 + 1)}} &&\nonumber\\\nonumber
&\times\begin{pmatrix}\mathrm{l_2} & \mathrm{k} & \mathrm{{l}'} \\ \mathrm{0} & \mathrm{0} & \mathrm{0}\end{pmatrix}
\begin{pmatrix}\mathrm{l_2} & \mathrm{k} & \mathrm{{l}'} \\ \mathrm{-m_2} & \mathrm{q} & \mathrm{{m}'}\end{pmatrix} 
\begin{pmatrix}\mathrm{l_3} & \mathrm{k} & \mathrm{{l}} \\ \mathrm{0} & \mathrm{0} & \mathrm{0}\end{pmatrix}
\begin{pmatrix}\mathrm{l_3} &  \mathrm{k} & \mathrm{{l}} \\ \mathrm{-m_3} & \mathrm{q}  & \mathrm{{m}}\end{pmatrix}&& \\ 
&\mathrm{\times{(-1)}^{m_2+m_3} \int_0^\infty  \frac{r^{k}_{<}}{r^{k+1}_{>}} P^{i*}_{l_2 m_2}(\rp)P^\epsilon_{{l}'{m}'}(\rp)d\rp P^i_{l_3,m_3}(r).} &&
\label{eq:exch_full}
\end{flalign}
For numerical efficiency, $\mathrm{P^\epsilon_{{l}'{m}'}(r)}$ are obtained solving \eq{eqn:mat1} with the non-iterative method described in ref. \citen{demekhin11}.   
 Diatomic molecules have rotational symmetry around the molecular axis and thus m is a good quantum number. Therefore,  $\mathrm{m_2}$ and $\mathrm{m_3}$ are equal and have a fixed value for a bound orbital i. In addition, $\mathrm{m}$ and $\mathrm{m'}$ are equal 
 and have a fixed value determined by the symmetry of the continuum orbital. For the $\mathrm{N_2}$ diatomic molecule, we find that convergence of the continuum orbital is achieved when considering l up to 19 for the single center 
 expansion. As a result, for each energy  $\mathrm{\epsilon}$, \eq{eqn:mat1} has at the most 19 degenerate solutions.

\subsection{Photo-ionisation Cross-sections}

The photo-ionisation cross-section\cite{ModernQM}  for an electron transitioning from an initial molecular orbital $\mathrm{\phi_{\rm i}}$ to a final continuum molecular orbital $\mathrm{\phi_{\epsilon}}$ is given by  
 \begin{flalign}
\mathrm{\sigma_{i\rightarrow \epsilon \rm} = \frac{4}{3}\alpha\pi^2 \omega N_i \sum_{M=-1,0,1} {|D_{i\epsilon}^M|}^{2},}&&
\end{flalign}
where $\mathrm{\alpha}$ is the fine structure constant, $\mathrm{N_i}$ is the occupation number of the initial molecular orbital i, $\mathrm{\omega}$ is the photon energy, and M is the polarization of the photon. In the length gauge, the matrix element $\mathrm{D_{i\epsilon}^M}$ is given by
 \begin{flalign}
 \label{cross}
\mathrm{D_{i\epsilon}^M = \int\phi_i({\bf r})\phi_\epsilon({\bf r})\sqrt{\frac{4\pi}{3}}rY_{1M}(\theta,\phi)d{\bf r}.}&&
\end{flalign}
In the single centre expansion formalism \eq{cross} takes the form
\begin{widetext}
\begin{flalign}
\label{eq:pi_wig}
\mathrm{D_{i\epsilon}^M}& = \mathrm{\sqrt{\frac{4\pi}{3}}\sum_{lm,l'm'}\int^\infty_0 dr P^{i*}_{lm}(r) r P^\epsilon_{l'm'}(r) \int d\Omega Y^{*}_{lm}(\theta,\phi)Y_{l'm'}(\theta,\phi)Y_{1M}(\theta,\phi)}&&\\\nonumber
 &\mathrm{ = \sum_{lm,l'm'}{(-1)}^{m} \sqrt{(2l+1)(2l'+1)}} \begin{pmatrix} 
\mathrm{l} & \mathrm{l'} & \mathrm{1} \\ 
\mathrm{0} & \mathrm{0} & \mathrm{0}
\end{pmatrix} \begin{pmatrix}
\mathrm{l} & \mathrm{l'} & \mathrm{1} \\ 
\mathrm{-m} & \mathrm{m'} & \mathrm{M}
\end{pmatrix} \mathrm{\int^\infty_0 dr P^{i*}_{lm}(r) r P^\epsilon_{l'm'}(r).}&&
\end{flalign}
\end{widetext}
In Table \ref{N2comp}, for certain transitions in $\mathrm{N_2}$, we compare with past calculations\cite{Semenov2000} the photo-ionization cross sections we compute using \eq{eq:pi_wig}. Very good agreement is obtained.
\begin{table}
\small
\begin{center}
 \begin{tabular*}{0.5\textwidth}{@{\extracolsep{\fill}}ccccc}
 \hline
$\mathrm{E_{photon}}$(eV) &  \multicolumn{2}{c}{$\mathrm{2\sigma_g\rightarrow \epsilon\sigma_u}$} & \multicolumn{2}{c}{$\mathrm{2\sigma_g\rightarrow \epsilon\pi_u}$}\    \\
 & Ref. \citen{Semenov2000} & This Work & Ref. \citen{Semenov2000} & This Work \\ \hline
40 & 0.035 & 0.038 & 0.073 & 0.20 \\ 
45 & 0.58 & 0.55 & 0.22 & 0.38 \\ 
50 & 2.6 & 2.5 & 0.41 & 0.52 \\ 
55 & 1.9 & 2.0 & 0.59 & 0.63 \\ 
60 & 1.1 & 1.2 & 0.71 & 0.69 \\ 
65 & 0.74 & 0.80 & 0.75 & 0.73 \\ 
70 & 0.54 & 0.58 & 0.74 & 0.73 \\ 
75 & 0.40 & 0.44 & 0.70 & 0.72 \\ 
80 & 0.31 & 0.33 & 0.65 & 0.69 \\
\hline
\end{tabular*}
\caption{Photo-ionization cross-sections for $\mathrm{N_2}$ transitions: columns 3 and 5 correspond to our results and  columns 2 and 4 correspond to previous calculations\cite{Semenov2000}}
\label{N2comp}
\end{center}
\end{table}

\subsection{Auger Rates}
In general, the Auger rate is given by\cite{FermiGR}
 \begin{flalign}
\mathrm{\Gamma=\overline{\sum}2\pi |\mathcal{M}|^2\equiv\overline{\sum}2\pi |\langle\Psi_{fin}|H_I|\Psi_{init}\rangle|^2,}&&
\label{GeneralAuger}
\end{flalign}
where $\mathrm{\overline{\sum}}$ denotes a summation over the final states and an average over the initial states. $\mathrm{|\Psi\rangle}$ is the wavefunction of all electrons in the molecular state. In the Hartree-Fock approximation, $\mathrm{|\Psi\rangle}$ is given by a Slater determinant of one-electron spin-orbital wavefunctions. The Auger transition is treated as a two-electron process and therefore the relevant part of $\mathrm{H_I}$ is the electron-electron Coulomb interaction term. In the second quantization formalism,\cite{Manne1985,InhesterThesis} this Hamiltonian term  is given by
\begin{flalign}
\mathrm{H_I^{ee}=\frac{1}{2}\sum_{\alpha\beta\gamma\delta}c_\alpha^\dag c_\beta^\dag c_\gamma c_\delta\langle \alpha\beta|\frac{1}{r_{12}}|\gamma\delta\rangle,}&&
\label{Hamiltonian}
\end{flalign}
where $\mathrm{c_\gamma}$ is the annihilation operator of the one-electron spin-orbital wavefunction $\mathrm{|\gamma\rangle}$ and $\mathrm{c^\dag_\alpha}$ is the creation operator of the one-electron spin-orbital wavefunction $\mathrm{|\alpha\rangle}$.
Then, the matrix element  takes the form

\begin{flalign}
\mathrm{\langle \Psi_{fin}|H_I^{ee}|\Psi_{init}\rangle_{Auger}=\frac{1}{2}\sum_{\alpha\beta\gamma\delta}\langle \Psi_{fin}|c_\alpha^\dag c_\beta^\dag c_\gamma c_\delta|\Psi_{init}\rangle\langle \alpha\beta|\frac{1}{r_{12}}|\gamma\delta\rangle.}&&
\label{CAO2}
\end{flalign}
 Using the anti-commutation relations of the creation and annihilation operators, \eq{CAO2} is written as
\begin{flalign}
\label{CAO13}
\mathrm{\langle \Psi_{fin}|H_I^{ee}|\Psi_{init}\rangle_{Auger}=\langle \zeta s|\frac{1}{r_{12}}|ba\rangle-\langle \zeta s|\frac{1}{r_{12}}|ab\rangle.}&&
\end{flalign}
In the Auger transition matrix element, a and b are the one-electron wavefunctions of the two valence electrons, while s and $\zeta$ are the one-electron wavefunctions of the core and continuum electrons, respectively. In this transition, the core hole in spin-orbital s is filled by an electron from spin-orbitals a or b, while  the other valence electron ionizes. Changing from the $\mathrm{m_a\mu_am_b\mu_b}$ to the $\mathrm{m_am_bSM_S}$ scheme we obtain 
\begin{flalign}
\label{SM_S4}
\mathrm{\mathcal{M}}&=\mathrm{\langle \Psi_{fin}|H_I^{ee}|\Psi_{init}\rangle_{Auger}}&&\nonumber\\
&=\mathrm{\delta_{S',S}\delta_{M'_S,M_S}\left(( \zeta s|\frac{1}{r_{12}}|ba)+(-1)^{S}( \zeta s|\frac{1}{r_{12}}|ab)\right).}&&
\end{flalign}
$\mathrm{m_a}$ and $\mathrm{\mu_a}$ are the projection of the orbital angular momentum and spin, respectively, while S is the total spin and $\mathrm{M_S}$ is the projection of the total spin. $\mathrm{( \zeta s|\frac{1}{r_{12}}|ba)}$ is the spatial part of the matrix element $\mathrm{\langle \zeta s|\frac{1}{r_{12}}|ba\rangle}$ which is given by
\begin{flalign}
\mathrm{( \zeta s|\frac{1}{r_{12}}|ba)=\int d{\bf r}\phi^*_\zeta\phi^*_s\frac{1}{r_{12}}\phi_b\phi_a.}&&
\label{SM_S5}
\end{flalign}
Using \eq{CAO13} and \eq{SM_S5} and expressing the orbital wavefunctions in the SCE scheme, we find 
\begin{flalign}
\label{AugerSM_S}
&\mathrm{\mathcal{M}=\delta_{S,S'}\delta_{M_S,M'_S}}&&\\\nonumber
&\mathrm{\times\left(\sum_{\substack{kl_\zeta l_s\\l_bl_a}}\sum_{q=-k}^{k}\int dr_1\int dr_2 P^{\zeta*}_{l_\zeta m_\zeta}(r_1)P^{s*}_{l_sm_s}(r_2)\frac{r^k_<}{r^{k+1}_>}P^{b}_{l_bm_b}(r_1)P^{a}_{l_am_a}(r_2)\right.}&&\\\nonumber
&\mathrm{\times(-1)^{m_s}\sqrt{(2l_s+1)(2l_a+1)}\gj{\mathrm{l_s}}{\mathrm{k}}{\mathrm{l_a}}{0}{0}{0}\gj{\mathrm{l_s}}{\mathrm{k}}{\mathrm{l_a}}{\mathrm{-m_s}}{\mathrm{q}}{\mathrm{m_a}}}&&\\\nonumber
&\mathrm{\mathrm{\times(-1)^{q+m_\zeta}\sqrt{(2l_\zeta +1)(2l_b+1)}\gj{\mathrm{k}}{\mathrm{l_\zeta} }{\mathrm{l_b}}{0}{0}{0}\gj{\mathrm{k}}{\mathrm{l_\zeta} }{\mathrm{l_b}}{\mathrm{-q}}{\mathrm{-m_\zeta}}{\mathrm{m_b}}}}&&\\\nonumber
&\mathrm{+(-1)^{S}\sum_{\substack{kl_\zeta l_s\\l_bl_a}}\sum_{q=-k}^{k}\int dr_1\int dr_2 P^{\zeta*}_{l_\zeta m_\zeta}(r_1)P^{s*}_{l_sm_s}(r_2)\frac{r^k_<}{r^{k+1}_>}P^{a}_{l_am_a}(r_1)P^{b}_{l_bm_b}(r_2)}&&\\\nonumber
&\mathrm{\times(-1)^{m_s}\sqrt{(2l_s+1)(2l_b+1)}\gj{\mathrm{l_s}}{\mathrm{k}}{\mathrm{l_b}}{0}{0}{0}\gj{\mathrm{l_s}}{\mathrm{k}}{\mathrm{l_b}}{\mathrm{-m_s}}{\mathrm{q}}{\mathrm{m_b}}}&&\\\nonumber
&\mathrm{\left.\times(-1)^{q+m_\zeta}\sqrt{(2l_\zeta +1)(2l_a+1)}\gj{\mathrm{k}}{\mathrm{l_\zeta} }{\mathrm{l_a}}{0}{0}{0}\gj{\mathrm{k}}{\mathrm{l_\zeta} }{\mathrm{l_a}}{\mathrm{-q}}{\mathrm{-m_\zeta}}{\mathrm{m_a}}\vphantom{\sum_{\substack{kl_\zeta l_s\\l_bl_a}}}\right)}&&
\end{flalign}
where $\mathrm{r_<=min(r,r')}$ and $\mathrm{r_>=max(r,r')}$. The Auger rate is given by
\begin{flalign}
\mathrm{\Gamma_{b,a\rightarrow s,\zeta}=
\sum_{\substack{\mathrm{m_am_bm_sm_\zeta}\\\mathrm{SM_SS'M'_S}}}\pi N_{12}N_h\sum_L|\mathcal{M}|^2},&&
\label{TotalSM_Sgeneral}
\end{flalign}
with $\mathrm{N_h}$  the number of holes in the orbital to be filled. $\mathrm{N_{12}}$ is the weighting occupation factor given by
 \begin{flalign}
\mathrm{N_{12}}=\Bigg\{\begin{array}{lr}
\dfrac{\mathrm{N_{v1}N_{v2}}}{\mathrm{2\times2}}\qquad \qquad & \mathrm{for\,\,different\,\, orbitals}\\[2ex]
\dfrac{\mathrm{N_{v1}(N_{v1}-1)}}{\mathrm{2\times2\times1}} &\mathrm{for\,\,same\,\, orbital}
\end{array}&&
\end{flalign}
where $\mathrm{N_{v1}}$ and $\mathrm{N_{v2}}$ are the occupations numbers of the valence orbitals that are involved in the Auger transition. Next, we compare our results for the Auger rates of  $\mathrm{N_2^+}$ with a 1s core-hole, which are computed using \eq{TotalSM_Sgeneral} with the Auger rates calculated using a Green's function method.\cite{Liegener} The 1s state corresponds to $\mathrm{\phi_{1s}=\frac{1}{\sqrt{2}}\left(\phi_{1\sigma_g}+\phi_{1\sigma_u}\right)}$. Using the orthogonality of the molecular states, it follows that the 1s Auger rates are obtained by averaging the Auger  rates of the $\mathrm{1\sigma_g}$ and $\mathrm{1\sigma_u}$ core hole molecular states. In the work in ref. \citen{Liegener}, only relative values of the Auger rates are given. Specifically, the ratio of each Auger rate with respect to the  transition is given. To compare the results in ref. \citen{Liegener} with our values we divide each Auger transition  by the sum of all Auger transitions for a 1s core-hole state. The resulting values are shown in Table \ref{Tab:AugerRate} and the agreement is shown to be good.  Moreover, we find that the sum of all Auger rates corresponding to a 1s core-hole is equal to $\mathrm{2.87E-3}$ a.u.. This value  compares well with the experimental value  of $\mathrm{3.77E-3}$ a.u. obtained in ref. \citen{Fang}. We note that the Auger rates we use to solve the rate equations are summed over all allowed spin configurations. The reason for this is that spin is not specified in the electronic configurations of the molecular states. 

\begin{table}
\small
\begin{center}
\begin{tabular*}{0.5\textwidth}{@{\extracolsep{\fill}}ccccc}
 \hline
Final State & Valence 1 & Valence 2 & This Work & Ref. \citen{Liegener} \\ \hline
$^3\mathrm{\Sigma_u^+}$ & 2$\mathrm{\sigma_u}$ & 3$\mathrm{\sigma_g}$ & 0.01 & 0.01 \\
$^1\mathrm{\Sigma_u^+}$ & 2$\mathrm{\sigma_u}$ & 3$\mathrm{\sigma_g}$ & 0.05 & 0.11 \\
$^3\mathrm{\Pi_u}$ & 3$\mathrm{\sigma_g}$ & 1$\mathrm{\pi_{ux}(1\pi_{uy})}$ & 0.01 & 0.01 \\
$^1\mathrm{\Pi_u}$ & 3$\mathrm{\sigma_g}$ & 1$\mathrm{\pi_{ux}(1\pi_{uy})}$ & 0.11 & 0.13 \\
$^3\mathrm{\Pi_g}$ & 2$\mathrm{\sigma_u}$ & 1$\mathrm{\pi_{ux}(1\pi_{uy})}$ & 0.02 & 0.03 \\
$^1\mathrm{\Pi_g}$ & 2$\mathrm{\sigma_u}$ & 1$\mathrm{\pi_{ux}(1\pi_{uy})}$ & 0.08 & 0.13 \\
$^3\mathrm{\Pi_u}$ & 2$\mathrm{\sigma_g}$ & 1$\mathrm{\pi_{ux}(1\pi_{uy})}$ & 0.03 & 0.01 \\
$^1\mathrm{\Pi_u}$ & 2$\mathrm{\sigma_g}$ & 1$\mathrm{\pi_{ux}(1\pi_{uy})}$ & 0.09 & 0.06 \\
$^3\mathrm{\Sigma_u^+}$ & 2$\mathrm{\sigma_g}$ & 2$\mathrm{\sigma_u}$ & 0.01 & 0.01 \\
$^1\mathrm{\Sigma_u^+}$ & 2$\mathrm{\sigma_g}$ & 2$\mathrm{\sigma_u}$ & 0.20 & 0.11 \\
$^3\mathrm{\Sigma_g^+}$ & 2$\mathrm{\sigma_g}$ & 3$\mathrm{\sigma_g}$ & 0.02 & 0.02 \\
$^1\mathrm{\Sigma_g^+}$ & 2$\mathrm{\sigma_g}$ & 3$\mathrm{\sigma_g}$ & 0.08 & 0.07 \\
$^1\mathrm{\Sigma_g^+}$ & 3$\mathrm{\sigma_g}$ & 3$\mathrm{\sigma_g}$ & 0.05 & 0.04 \\
$^1\mathrm{\Delta_g}$ & 1$\mathrm{\pi_{ux}(1\pi_{uy})}$ & 1$\mathrm{\pi_{ux}(1\pi_{uy})}$ & 0.09 & 0.12 \\
$^1\mathrm{\Sigma_g^+}$ & 1$\mathrm{\pi_{ux}(1\pi_{uy})}$ & 1$\mathrm{\pi_{uy}(1\pi_{ux})}$ & 0.03 & 0.01 \\
$^1\mathrm{\Sigma_g^+}$ & 2$\mathrm{\sigma_u}$ & 2$\mathrm{\sigma_u}$ & 0.05 & 0.13 \\
$^1\mathrm{\Sigma_g^+}$ & 2$\mathrm{\sigma_g}$ & 2$\mathrm{\sigma_g}$ & 0.07 & 0.02 \\ \hline
\end{tabular*}
\caption{Ratio of each Auger transition for a 1s core-hole divided by the sum of all Auger transitions for a 1s core-hole for  $\mathrm{N_2}$}
\label{Tab:AugerRate}
\end{center}
\end{table}

\section{Results}
Using the methods described in the previous sections, we compute the photo-ionization cross sections and the Auger transitions for all allowed molecular transitions up to $\mathrm{N_{2}^{4+}}$ for 525 eV and 1100 eV FEL pulses. $\mathrm{N_{2}^{4+}}$ is the highest molecular ion state that can be reached, since  we assume that the  $\mathrm{N_{2}^{4+}}$ state  dissociates instantaneously. In addition, using the method we developed in ref. \citen{Wallis} we compute the Auger rates for all allowed atomic transitions while we obtain the atomic cross sections  from ref. \citen{LANL}. We solve 309 rate equations to obtain the ion yields and the electron spectra. In addition, in order to obtain the population of all accessible pathways, we solve roughly 1.8$\times 10^{6}$ rate equations for the 525 eV FEL pulse and 6.6$\times 10^{6}$ rate equations for the 1100 eV FEL pulse. The computation of the pathway population allows us to identify the percentage of the contribution of SCH versus DCH molecular states to the final atomic ion yields. 

\subsection{Ion Yields}

\begin{figure}
\centering
\includegraphics[width=\linewidth]{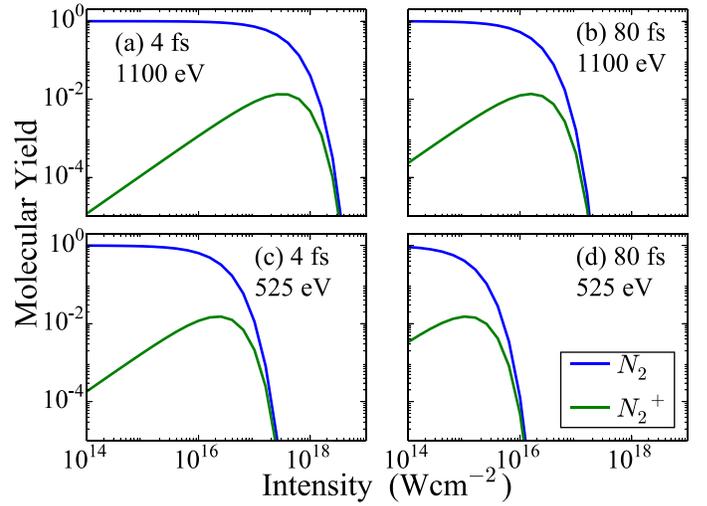}
\caption{Molecular ion yields resulting from the interaction of $\mathrm{N_{2}}$ with four FEL pulses as a function of the intensity of the laser pulse}
\label{fig:IYM4}
\end{figure}

First, we compute the molecular and atomic ion yields for a 525 eV and 1100 eV FEL pulses. For each photon energy we consider two different full width half maximum (FWHM) durations of the FEL pulse, namely 4 fs and 80 fs.  In Fig. \ref{fig:IYM4}, we show the dependence on intensity of the molecular ion yields resulting from the interaction of $\mathrm{N_{2}}$ with four different FEL pulses. We find that only the $\mathrm{N_2}$ and $\mathrm{N_2^+}$ states are populated a long time after the end of the FEL pulses. This is expected, since, in our model, all higher charged states eventually dissociate. We also find that after a certain  intensity the population of both the  $\mathrm{N_2}$ and the $\mathrm{N_2^+}$ molecular states  reduces significantly for all four FEL pulses considered.
For the same photon energy FEL pulses,  $\mathrm{N_2}$ and $\mathrm{N_2^+}$ are depleted at a smaller intensity for the 80 fs FEL-pulse compared to the  4 fs one. Comparing the molecular ion yields  of  FEL pulses with the same FWHM but different photon energy, we find that  depletion of $\mathrm{N_2}$ and $\mathrm{N_2^+}$ occurs at a smaller intensity for the 525 eV FEL pulse. The reason for this is that the single-photon ionization cross sections for the molecular transitions,   are larger for the 525 eV FEL pulse  compared to the 1100 eV FEL pulse, for a certain intensity.

\begin{figure*}[]
\centering
\includegraphics[width=0.82\linewidth]{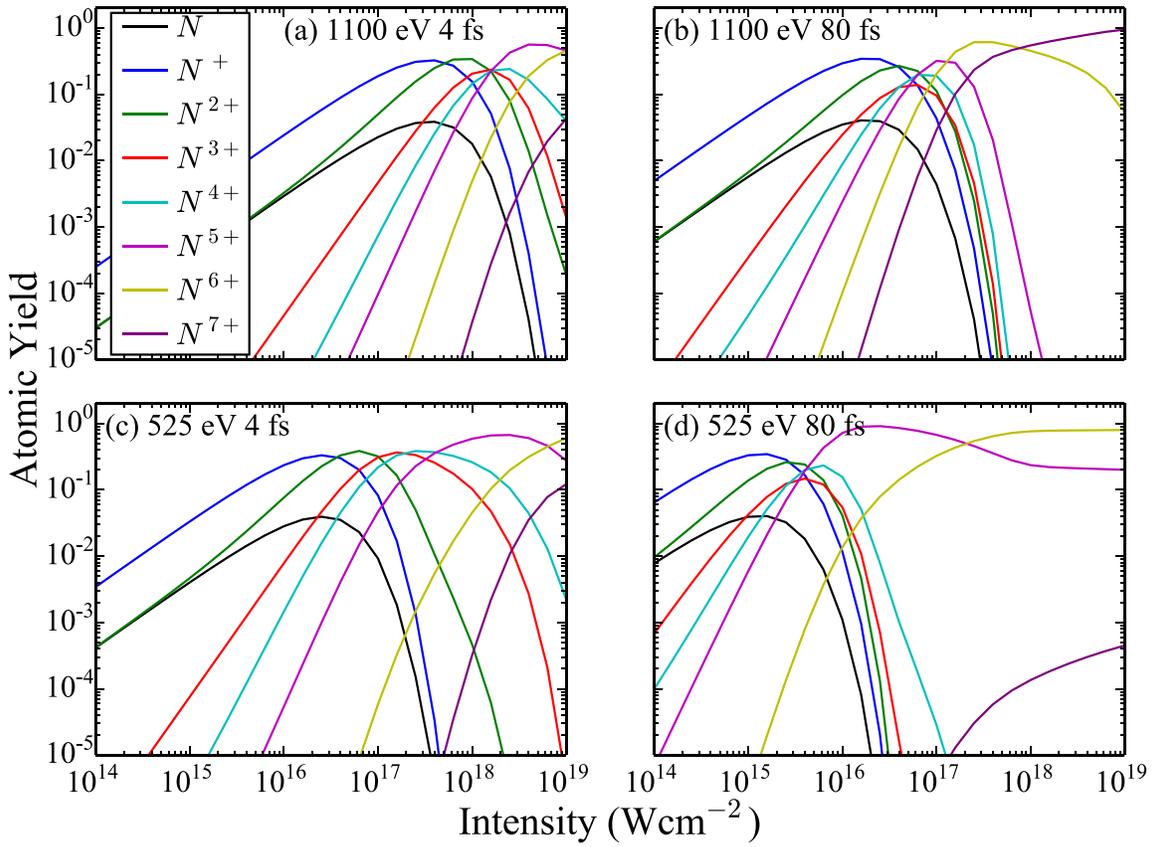}
\caption{Atomic ion yields resulting from the interaction of $\mathrm{N_{2}}$ with four FEL pulses as a function of the intensity of the laser pulse}
\label{fig:IYA4}
\end{figure*}

Next, in Fig. \ref{fig:IYA4} we show the atomic ion yields  for the same four FEL pulses as in  Fig. \ref{fig:IYM4} as a function of the laser intensity. Comparing the atomic ion yields  of  FEL pulses with the same photon energy  but different FWHM, we find that 
the population of the higher charged states becomes significant at smaller intensities for the 80 fs FEL pulse compared to the 4 fs one. This is expected, since more single-photon ionization processes take place during the longer FEL pulse. Next, 
 we compare the atomic ion yields  of  FEL pulses with the same  FWHM but different photon energy. We find that 
the population of the higher charged states becomes significant at smaller intensities for the 525 eV FEL pulse compared to the 1100 eV one. The reason for this is that the single-photon ionization cross sections both for the atomic and the molecular transitions are higher for the 525 eV FEL pulse. 

\subsection{Single versus Double Core Holes}
 \begin{figure}
\centering
\includegraphics[width=1\linewidth]{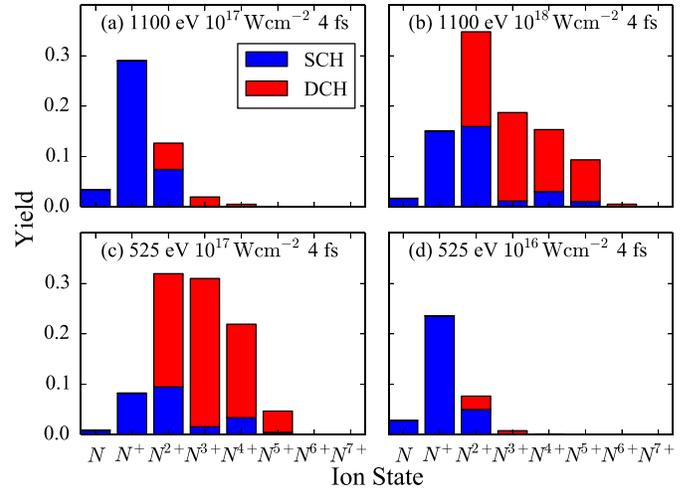}
\caption{Atomic ion yields and contribution of SCH and DCH molecular states for FEL pulses with  4 fs FWHM and photon energy of 525 eV and  of 1100 eV }
\label{fig:PathwaysBoth}
\end{figure}

We now compute all energetically accessible pathways that start from the $\mathrm{N_{2}}$ ground state and end at ion fragments up to $\mathrm{N^{7+}}$ for a 4 fs FEL pulse. We identify the contribution to the atomic ion yields of pathways that have accessed only a SCH molecular state versus pathways that have accessed a DCH molecular state. This contribution is shown in Fig. \ref{fig:PathwaysBoth} for a 525 eV and a 1100 eV  FEL pulse at different laser pulse intensities. 

First, we compare the results for two FEL pulses that have  the same intensity of $\mathrm{10^{17}\;Wcm^{-2}}$ but different photon energy, see Fig. \ref{fig:PathwaysBoth} (a) and (c). We find that the population of the higher charged atomic ion states is much larger for the 525 eV FEL pulse. The reason for this is that  the molecular and atomic photo-ionization cross sections  as well as the photon flux are larger for the 525 eV FEL pulse. Another consequence of the larger molecular photo-ionization cross sections is that for the 525 eV FEL pulse it is more probable for a second core hole to be created by single-photon ionization before an Auger transition takes place. This explains why the DCH molecular states contribute significantly more than the SCH molecular states to the population of the higher charged atomic ions, see Fig. \ref{fig:PathwaysBoth}(c). However, if a smaller intensity of $\mathrm{10^{16}\;Wcm^{-2}}$  is considered for the 525 eV FEL pulse, see  Fig. \ref{fig:PathwaysBoth}(d), then similar results are obtained as for the 1100 eV FEL pulse at an intensity of $\mathrm{10^{17}\;Wcm^{-2}}$, see Fig. \ref{fig:PathwaysBoth}(a).
In Fig. \ref{fig:PathwaysBoth}(b), it is shown that when a  higher intensity of  $\mathrm{10^{18}\;Wcm^{-2}}$ is considered for the 1100 eV FEL pulse,  similar results are obtained as for the 525 eV FEL at a smaller intensity, see Fig. \ref{fig:PathwaysBoth}(c). That is, higher charged atomic ion states are significantly populated and the DCH molecular states contribute significantly to these yields. The reason for this is that the higher intensity counteracts the effects from the single-photon ionization cross sections being smaller for the 1100 eV FEL pulse compared to the 525 eV FEL pulse.

\subsection{Electron spectra}
\begin{figure*}
 \centering
\includegraphics[width=0.82\linewidth]{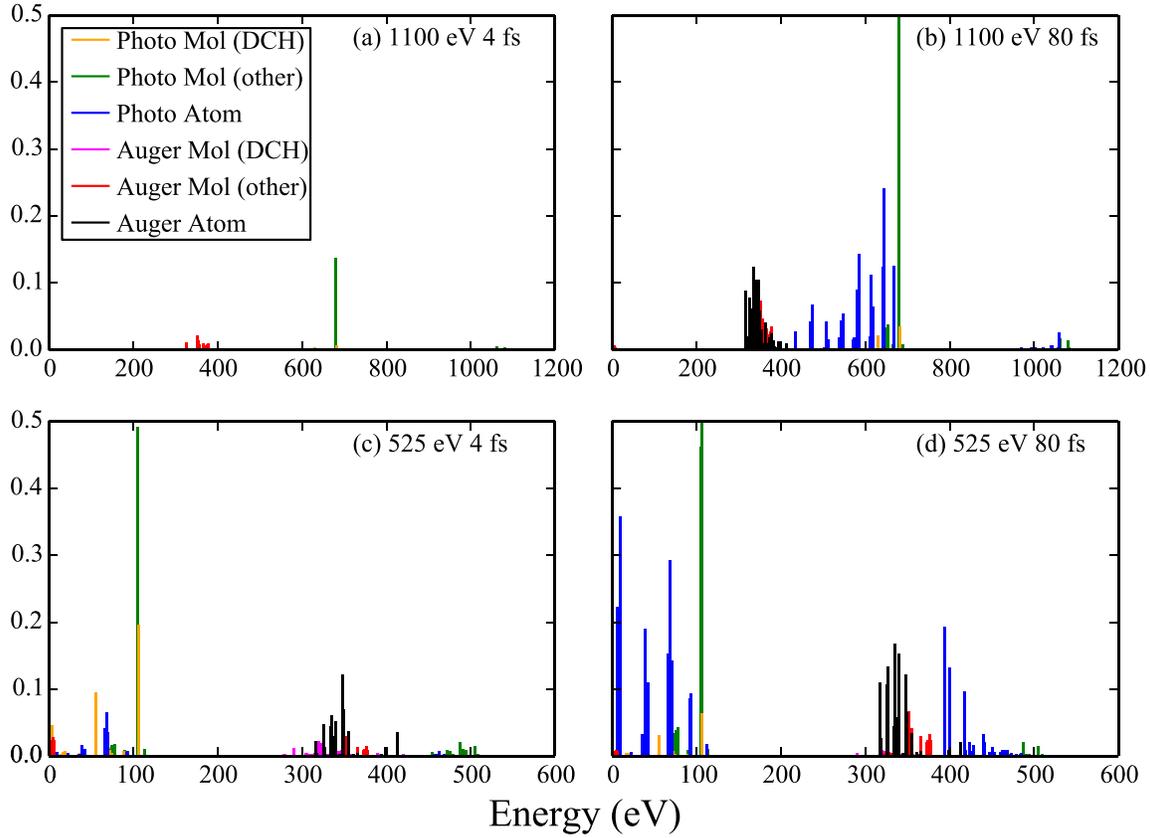}
\caption{Electron spectra resulting from the interaction of $\mathrm{N_{2}}$ with  FEL pulses at an intensity of $\mathrm{10^{17}\;Wcm^{-2}}$}
\label{fig:N2_ElecSpectra}
\end{figure*}

Using the molecular and atomic Auger and photo-ionization yields, we plot the probability for an electron to ionize with a certain energy, that is, we plot the electron spectra  in Fig. \ref{fig:N2_ElecSpectra}. Specifically, we plot the electron spectra for 525 eV and 1100 eV FEL pulses. For each photon energy two FWHM laser pulses are considered, namely of  4 fs and of 80 fs. Comparing the electron spectra corresponding to the same FWHM FEL pulse, i.e.  Fig. \ref{fig:N2_ElecSpectra} (a) with (c) and  Fig. \ref{fig:N2_ElecSpectra} (b) with (d), we find that 
there is a higher contribution of atomic transitions to the electron spectra  for the 525 eV FEL pulse. As previously mentioned, the reason is that the molecular photo-ionization cross sections are larger for the smaller photon energy resulting in a larger population reaching higher charged molecular states which in turn dissociate into atomic ion fragments. The atomic photo-ionization cross sections are also higher for the 525 eV FEL pulse and so higher charged atomic ion states are thus reached. Moreover, comparing 
the electron spectra corresponding to the same photon energy FEL pulse, i.e.  Fig. \ref{fig:N2_ElecSpectra} (a) with (b) and  Fig. \ref{fig:N2_ElecSpectra} (c) with (d) we find that there is a higher contribution of atomic transitions to the electron spectra for the longer duration, 80 fs, FEL pulse. The longer pulse duration allows for a larger number of photo-ionization processes to take place which in turn leads to the production of more higher charged atomic ion states. We also note that for the 525 eV FEL pulse the electrons can escape with small energies. Thus, it is important that in our formulation the Auger and the photo-ionization rates are computed using molecular and not atomic continuum orbitals.

Finally, we investigate whether studying electron spectra is an efficient way of detecting the formation of DCH molecular states. To answer this question we plot the contribution to the Auger and photo-ionization yields of DCH molecular states. We find  that the contribution to the electron spectra of DCH Auger molecular transitions is difficult to distinguish and overlaps with Auger atomic transitions. However, the DCH single-photon ionization molecular transitions  are easier to distinguish. For instance there is a clear peak in the electron spectra due to DCH single-photon ionization molecular transitions at around 55 eV for the case of the 525 eV 4 fs FEL pulse. This result suggests that photoionization electron spectra maybe a viable route for the detection of DCH molecular transitions.

\section{Conclusions}

We investigate the interaction of molecular nitrogen with FEL radiation. We computed molecular continuum orbitals in the single center expansion scheme and used these orbitals to compute  the Auger rates and photo-ionization cross-sections for molecular nitrogen. Formulating rate equations for all energetically accessible molecular and atomic transitions, we investigated the dependence of the final fragments yields on the parameters of the FEL pulse. Moreover, we studied the contribution of the DCH molecular states to the final atomic ion yields. We found that for a relatively small photon energy of 525 eV, with 420 eV being the photon energy needed to create a core hole, already at intermediate intensities,  DCH molecular states contribute significantly  to the formation of the final atomic ion fragments.  For a higher photon energy of 1100 eV FEL pulse, we find that a much higher intensity is needed in order for the DCH molecular states to significantly contribute to the final atomic ion fragments. Finally, we computed the contribution of the Auger and the single-photon ionization processes in the electron spectra. Our results suggest that single-photon ionization processes are a more efficient tool for detecting the formation of DCH molecular states in the electron spectra. Additional studies are needed to verify this and to investigate  the effect of nuclear motion.  

\section*{Acknowledgements}

 A. E. is grateful to Peter Lambropoulos for pointing out DCH formation as an interesting problem in molecules interacting with FEL pulses. A.E. acknowledges the  use of the Legion computational resources at UCL.





\appendix


\section{Direct and Exchange Coefficients} \label{App:aAndb}

In the Hartree-Fock framework, after applying the variational principle,\cite{BranJoa} the electron-electron interaction terms can be written as
\begin{flalign}
\mathrm{\sum_i^{orbs}a_iJ_{i}\phi_\epsilon-\sum_i^{orbs}b_iK_{i}\phi_\epsilon=\epsilon^{ee}\phi_\epsilon,}&&
\label{ContHF}
\end{flalign}
where $\mathrm{\phi_\epsilon}$ is the spin-orbital of the molecular continuum electron with spin orientation $\mathrm{\mu_\epsilon}$ and $\mathrm{\epsilon^{ee}}$ is the energy contribution of the electron-electron interaction terms. The index i refers to a bound molecular orbital and $\mathrm{J_i}$ and $\mathrm{K_i}$ are defined as
\begin{flalign}
\label{JKdefHF}
\mathrm{J_{i}\phi_\epsilon} &= \mathrm{\langle \phi_i|\frac{1}{r_{12}}|\phi_i\rangle \phi_\epsilon}&& \\\nonumber
\mathrm{K_i \phi_\epsilon} &= \mathrm{\langle \phi_i|\frac{1}{r_{12}}|\phi_\epsilon\rangle \phi_i.}&&
\end{flalign}
To obtain the $\mathrm{a_i}$ and $\mathrm{b_i}$ coefficients in the general case, it suffices to obtain $\mathrm{a_i}$ and $\mathrm{b_i}$ for three limiting cases. Since we consider molecular orbitals  in all three limiting cases, the electron occupancy of the shells involved is  zero, one or two. If a shell is not occupied, the coefficients $\mathrm{a_i}$ and $\mathrm{b_i}$ are zero. For the first  limiting case, a two-electron system is considered, with both electrons initially occupying a single shell i and one of these electrons finally being emitted to the continuum. Spin is conserved and it is equal to zero in the initial and final states. Therefore, a two-electron wavefunction must be constructed that is anti-symmetric in spin and anti-symmetric under exchange of electrons. Such a wavefunction is given as a sum of the following two Slater determinants
\begin{flalign}
\mathrm{\Phi(q_1,q_2)=}\frac{\mathrm{1}}{\mathrm{\sqrt{2}}}\left(\frac{\mathrm{1}}{\mathrm{\sqrt{2!}}}
\begin{vmatrix}\mathrm{\phi_i^\uparrow(q_1)}&\mathrm{\phi_\epsilon^\downarrow(q_1)}\\
\mathrm{\phi_i^\uparrow(q_2)}&\mathrm{\phi_\epsilon^\downarrow(q_2)}\end{vmatrix}
-\frac{\mathrm{1}}{\mathrm{\sqrt{2!}}}\begin{vmatrix}\mathrm{\phi_i^\downarrow(q_1)}&\mathrm{\phi_\epsilon^\uparrow(q_1)}\\
\mathrm{\phi_i^\downarrow(q_2)}&\mathrm{\phi_\epsilon^\uparrow(q_2)}\end{vmatrix}\right),&&
\label{SlaterDet2sing}
\end{flalign}
where $q_1$ and $q_2$ are the spin and space coordinates of the two electrons. Using spin conservation and exchange symmetry, it is found that the energy contribution of the electron-electron interaction term is given by
\begin{flalign}
\mathrm{\epsilon^{ee}=\langle\Phi|\frac{1}{r_{12}}|\Phi\rangle=\langle \phi_i\phi_\epsilon|\frac{1}{r_{12}}|\phi_i\phi_\epsilon\rangle+\langle \phi_i\phi_\epsilon|\frac{1}{r_{12}}|\phi_\epsilon\phi_i\rangle.}&&
\label{EnergySinglet}
\end{flalign}
Using the variational principle in the Hartree-Fock equations scheme\cite{BranJoa} for the continuum orbital, the following equations are obtained
\begin{flalign}
\mathrm{J_{i}\phi_\epsilon+K_{i}\phi_\epsilon=\epsilon^{ee}\phi_\epsilon.}&&
\label{singletHF}
\end{flalign}
Comparing \eq{ContHF} and \eq{singletHF}, we find that $\mathrm{a_i=1}$ and $\mathrm{b_i=-1}$.

Another limiting case involves two shells i and j. In the initial state one electron is in shell i and two electrons occupy shell j. In the final state one electron from the j shell escapes to the continuum. A three-electron wavefunction must be constructed which is anti-symmetric in spin regarding the continuum electron and the electron in the j shell and anti-symmetric under exchange of electrons. Such a wavefunction is given as a sum of the following two Slater determinants
\begin{flalign}
\label{SlaterDet3Closed}
\mathrm{\Phi(q_1,q_2,q_3)}&=\frac{\mathrm{1}}{\mathrm{\sqrt{2\times3!}}}
\begin{vmatrix}\mathrm{\phi_i^\uparrow(q_1)}&\mathrm{\phi_j^\downarrow(q_1)}&\mathrm{\phi_\epsilon^\uparrow(q_1)}\\
\mathrm{\phi_i^\uparrow(q_2)}&\mathrm{\phi_j^\downarrow(q_2)}&\mathrm{\phi_\epsilon^\uparrow(q_2)}\\
\mathrm{\phi_i^\uparrow(q_3)}&\mathrm{\phi_j^\downarrow(q_3)}&\mathrm{\phi_\epsilon^\uparrow(q_3)}\end{vmatrix}&&\\\nonumber
&-\frac{\mathrm{1}}{\mathrm{\sqrt{2\times3!}}}
\begin{vmatrix}\mathrm{\phi_i^\uparrow(q_1)}&\mathrm{\phi_j^\uparrow(q_1)}&\mathrm{\phi_\epsilon^\downarrow(q_1)}\\
\mathrm{\phi_i^\uparrow(q_2)}&\mathrm{\phi_j^\uparrow(q_2)}&\mathrm{\phi_\epsilon^\downarrow(q_2)}\\
\mathrm{\phi_i^\uparrow(q_3)}&\mathrm{\phi_j^\uparrow(q_3)}&\mathrm{\phi_\epsilon^\downarrow(q_3)}\end{vmatrix}.&&
\end{flalign}
Following the same procedure as for the other limiting case, the following equations are obtained
\begin{flalign}
\mathrm{\left(J_{i}-\tfrac{1}{2}K_{i}+J_{j}+K_{j}\right)\phi_\epsilon=\epsilon^{ee}\phi_\epsilon,}&&
\label{3HF}
\end{flalign}
Comparing \eq{ContHF} and \eq{3HF}, it is found that $\mathrm{a_j=1}$ and $\mathrm{b_j=-1}$, while $\mathrm{a_i=1}$ and $\mathrm{b_i=\tfrac{1}{2}}$.

The third limiting case involves two electrons occupying shell i and two electrons occupying shell j in the initial state, with one electron from orbital j escaping to the continuum in the final state. Following the same procedure as in the other two cases, it can be shown that $\mathrm{a_i=2}$ and $\mathrm{b_i=1}$ and $\mathrm{a_j=1}$ and $\mathrm{b_j=-1}$. In general, for all molecular ion states, in the Hartree-Fock formalism, the occupation coefficients $\mathrm{a_i}$ and $\mathrm{b_i}$ can be obtained using the above three limiting cases.

\bibliography{Auger} 
\bibliographystyle{rsc} 

\end{document}